\documentclass[12pt,journal,a4paper,draftcls,oneside,onecolumn]{IEEEtran}
\usepackage{times} 
\usepackage{graphicx}
\usepackage{cite}
\usepackage{citesort}
\usepackage{color}
\usepackage{psfrag}
\usepackage{subfigure}
\usepackage{amssymb}
\usepackage{epsfig}
\usepackage{pifont}
\usepackage{amsmath}
\usepackage{array}
\usepackage{multicol}
\usepackage[T1]{fontenc}
\usepackage[ruled, lined, linesnumbered, commentsnumbered, longend]{algorithm2e}
\usepackage{algorithmic}
\usepackage{comment}

\renewcommand{\baselinestretch}{1.2} %for Lie-Liang to work
\makeatletter
\def\ifundefined{\@ifundefined}
\makeatother 

\begin{document}

%\vspace*{1cm}

\title{{Optimization for MIMO Integrated Sensing and
    Communications}
\author{Lie-Liang Yang}
\thanks{L.-L. Yang is with the School of Electronics and Computer Science, University of Southampton, SO17 1BJ, UK. (E-mail: lly@ecs.soton.ac.uk). This document is a chapter in the book: L.-L. Yang, J. Shi, K.-T. Feng, L.-H. Shen, S.-H. Wu and T.-S. Lee, Resource Optimization in Wireless Communications: Fundamentals, Algorithms and Applications, Academic Press, USA (to be published in 2024).}}

%\thanks{This work has been funded .}}

\maketitle

\begin{abstract}
The fundamentals of MIMO communications and MIMO sensing are firstly
analyzed with regard to channel and sensing capacities. It is shown
that the different objectives of communications and sensing lead to
different signaling waveforms required for achieving their
capacities. Hence, the optimization of integrated sensing and
communications (ISAC) is relied on a trade-off expected between the
performance of communications and that of sensing. Following this
observation, the design and resource optimization in general MIMO ISAC
systems are discussed along with the analysis of some existing ISAC
schemes. Furthermore, the design of ISAC in mmWave communications is
addressed. Specifically, the principle of sensing in mmWave systems is
established, and a range of optimization alternatives for ISAC design
in mmWave systems are reviewed.
\end{abstract}


%%%
\section{Introduction}\label{subsection-6G-6.1x}
%%%

Wireless signals are sent mainly for one of the two applications, namely  communication and sensing\index{Sensing!wireless}, which are referred to as wireless communications and wireless sensing, respectively.  Wireless communications and wireless sensing have their respective objectives. In wireless communications, the objective is to transmit information, preferably to transmit at the highest possible information rate over certain geographic distance for given resources, such as frequency band and power, available. By contrast, in wireless sensing systems, the objective is to acquire the knowledge about environment. For given resources, also such as frequency band and power, available, the objective is to acquire the knowledge about environment as more as possible and as accurate as possible. Hence, both  wireless communications systems and wireless sensing systems make use of the similar resources, but their served objectives are very different, resulting in that the signaling for communications and that for sensing typically differently. Nevertheless, considering the limited and costly resources from radio spectrum, it would be highly desirable if the same radio resources can be shared by the communications and sensing systems to achieve their different objectives, while imposing little trade-off between their performance. Furthermore, it would be beneficial if the cooperation between communication and sensing functions can enable to improve efficiency, as well as to reduce device size, cost and power consumption. Towards these objectives, integrated sensing and communication (ISAC) addresses the theories and corresponding technologies, enabling to integrate the communication and sensing functionalities, so as to make efficient use of resources and to mutually benefit each other~\cite{9737357}.

\begin{figure}[tb]
  \begin{center}
    \includegraphics[width=0.98\linewidth]{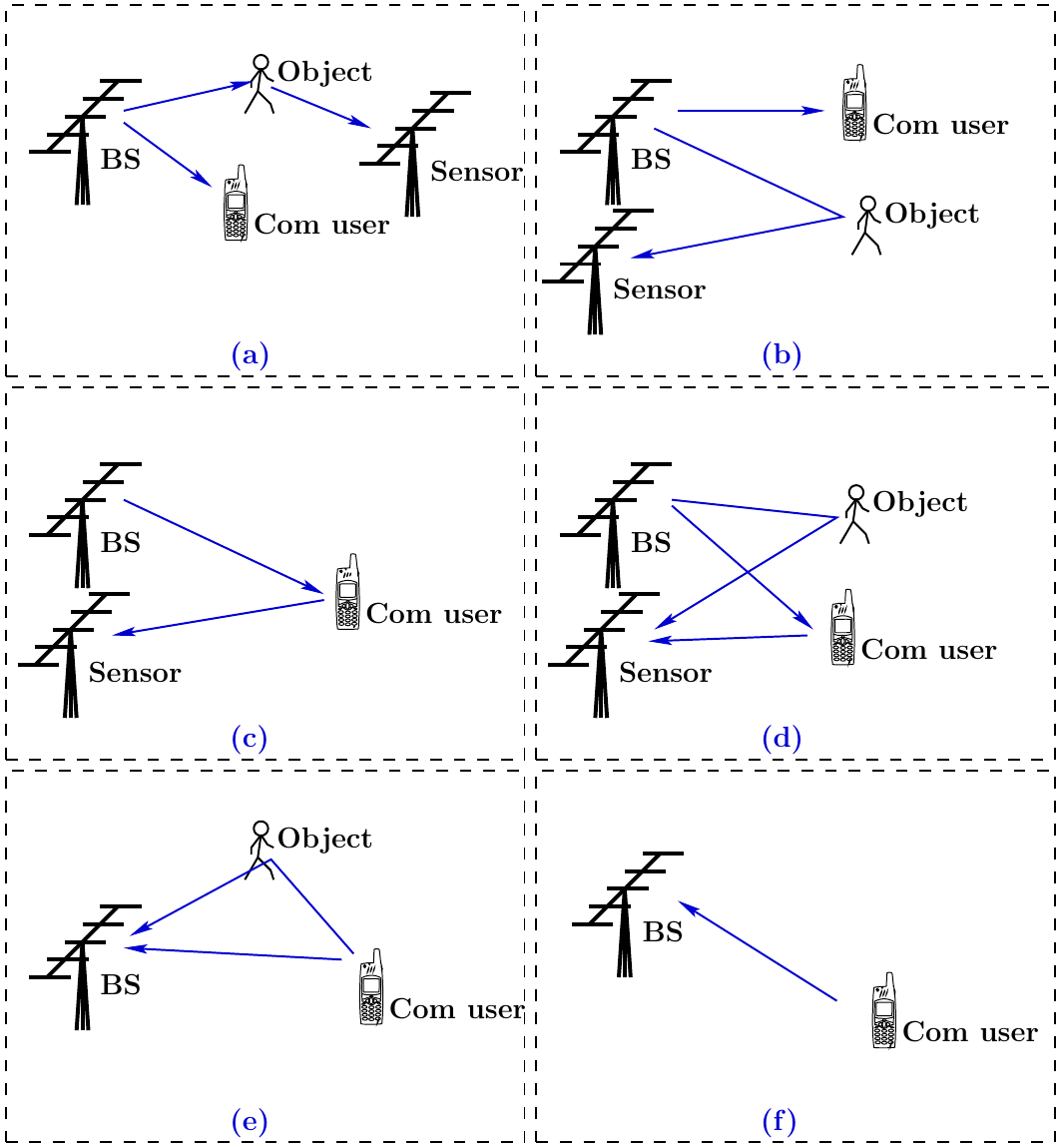}
 
  \end{center}
  \caption{Scenarios for implementation of ISAC.}
\label{figure-ISAC-scenarios}
\end{figure}

In ISAC systems, there are both communication and sensing functions
accomplished by different terminals (devices) involved. While
communication implements information transmission from a transmitter
to a receiver, the sensing function may be implemented at different
terminals and at different locations.
Fig.~\ref{figure-ISAC-scenarios} shows some ISAC scenarios. To be more
specific, corresponding to the uplink and downlink in cellular
communications systems, in the cellular-style ISAC systems, sensing
can be classified into downlink sensing\index{Sensing!downlink}, as
shown in Figs.~\ref{figure-ISAC-scenarios} (a)-(d) and uplink
sensing\index{Sensing!uplink}, as shown in
Figs.~\ref{figure-ISAC-scenarios} (e)-(f)~\cite{8827589}. In uplink
sensing, BS uses the uplink communication signals sent from mobile
users to sense the locations, velocities, etc., of the mobile users
themselves, as seen in Fig.~\ref{figure-ISAC-scenarios} (f), or of the
scatters in the environment, Fig.~\ref{figure-ISAC-scenarios} (e). By
contrast, in downlink sensing, the signals used for sensing are sent
by BS, as seen in Figs.~\ref{figure-ISAC-scenarios} (a)-(d).  Downlink
sensing can also be divided into downlink active
sensing\index{Sensing!downlink active}, as illustrated in
Figs.~\ref{figure-ISAC-scenarios} (b)-(d), and downlink passive
sensing\index{Sensing!downlink passive}, as depicted in
Fig.~\ref{figure-ISAC-scenarios} (a).  With the downlink passive
sensing, a sensing terminal uses the signals sent from BS to sense the
objects or mobile users in the environments. In the downlink active
sensing, the antenna array used for sensing is co-located with the
antenna array of BS or is simply the BS antenna array. To sense the
objects in the environment or/and the mobile users, BS first sends
probing signals, which may simply be the communication signals, and
then uses the echo signals to derive the sensing results as required.

In this chapter, the fundamentals of MIMO communications and MIMO sensing are first analyzed in Sections~\ref{subsection-6G-6.2.1} and \ref{subsection-6G-6.2.2}, respectively, to demonstrate their differences in terms of design and optimization and hence, to explain the challenges of design and optimization in ISAC systems. In Section~\ref{subsection-6G-6.3}, several design and resource optimization issues about MIMO ISAC systems are discussed. Finally, in Section~\ref{subsection-6G-6.4}, the ISAC design and optimization in mmWave ISAC systems are explored, when several exemplified cases are considered.      

\section{Fundamentals of MIMO Communications and MIMO Sensing}\label{subsection-6G-6.2}

%%%%%%%
\subsection{MIMO Communications}\label{subsection-6G-6.2.1}\index{MIMO!communications|(}
%%%%%%%

%
\begin{figure}[tb]
  \begin{center}
    \includegraphics[width=0.75\linewidth]{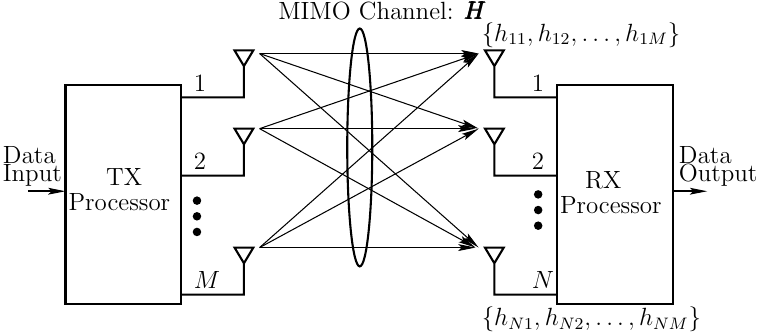}
 
  \end{center}
  \caption{Diagram of MIMO systems.}
\label{figure-MIMO-frame}
\end{figure}

Let us consider a MIMO system as shown by
Fig.~\ref{figure-MIMO-frame}, where transmitter and receiver are
equipped with $M$ and $N$ antennas, respectively.  Assume that the
transmitted signals over $T$ successive transmissions are
$\pmb{X}=[\pmb{x}_1,\pmb{x}_2,\ldots,\pmb{x}_T]$, where
$\pmb{X}\in\mathbb{C}^{M\times T}$.  Then, the received signals by the $N$
receive antennas corresponding to the $T$ transmissions can be written
as
\begin{align}\label{eq:ISAC-1}
\pmb{Y}=\pmb{H}\pmb{X}+\pmb{N}
\end{align}
where $\pmb{Y}\in\mathbb{C}^{N\times T}$ are observations at receiver,
$\pmb{N}\in\mathbb{C}^{N\times T}$ are complex Gaussian noise matrix
of each element independently distributed with zero mean and a
variance of $\sigma^2/2$ per dimension, $\pmb{H}\in\mathbb{C}^{N\times
  M}$ is the MIMO channel matrix connecting the transmit and receive
antennas.

For optimal communication design, given the total transmit power and
the channel knowledge of $\pmb{H}$ known to both transmitter and
receiver, the objective is to design $\pmb{X}$ that satisfies certain
characteristics (to be determined below), so that the mutual
information between the transmitted $\pmb{X}$ and the received
observation $\pmb{Y}$ is maximum. The $\pmb{X}$ designed in such way
will enable the MIMO system to achieve the capacity counted by the
number of bits conveyed per second per Hertz (b/s/Hz). According to
\cite{book:Thomas-Cover-Information-Theory}, the mutual
information\index{Mutual information} between $\pmb{X}$ and $\pmb{Y}$
conditioned on $\pmb{H}$ can be formulated as
\begin{align}\label{eq:ISAC-2}
I(\pmb{X};\pmb{Y}|\pmb{H})=&h(\pmb{Y}|\pmb{H})-h(\pmb{Y}|\pmb{X},\pmb{H})\nonumber\\
=&h(\pmb{Y}|\pmb{H})-h(\pmb{N})
\end{align}
where $h(\cdot)$ represents the differential
entropy\index{Differential entropy} of the corresponding argument. In
\eqref{eq:ISAC-2}, $h(\pmb{N})$ is the differential entropy of the
Gaussian noise matrix, which can be derived from the Gaussian
distribution of $\pmb{N}$ and is given by\footnote{More details can be
  found below in Section~\ref{subsection-6G-6.2.2}.}
\begin{align}\label{eq:ISAC-3}
h(\pmb{N})=T\log_2\left[\left(\pi e\right)^{N}\det(\sigma^2\pmb{I}_N)\right]
\end{align}
where $\det(\pmb{A})$ denotes the determinant of a square matrix
$\pmb{A}$. The differential entropy of $\pmb{Y}$ given $\pmb{H}$ is given by
\begin{align}\label{eq:ISAC-4}
h(\pmb{Y}|\pmb{H})=\log_2\left[\left(\pi e\right)^{NT}\det(\pmb{R}_y)\right]
\end{align}
where $\pmb{R}_y$ is the auto-correlation matrix of $\pmb{Y}$. When
expressing $\pmb{Y}=\left[\pmb{y}_1,\pmb{y}_2,\ldots,\pmb{y}_T\right]$, it
can be readily show that
\begin{align}\label{eq:ISAC-5}
\pmb{R}_y=E\left[\pmb{Y}\pmb{Y}^H\right]=E\left[\sum_{t=1}^T\pmb{y}_t\pmb{y}_t^H\right]=T\pmb{R}_{\pmb{y}_t}
\end{align}
where $\pmb{R}_{\pmb{y}_t}$ is the auto-correlation matrix of the
columns of $\pmb{Y}$, i.e., 
\begin{align}\label{eq:ISAC-6}
\pmb{R}_{\pmb{y}_t}=E\left[\pmb{y}_t\pmb{y}_t^H\right]=\pmb{H}\pmb{Q}_x\pmb{H}^H+\sigma^2\pmb{I}_N
\end{align}
where $\pmb{Q}_x=E\left[\pmb{x}_t\pmb{x}_t^H\right]\in\mathbb{C}^{M\times M}$ is the covariance
matrix of the transmitted signals $\pmb{x}_t$.
 
Upon substituting \eqref{eq:ISAC-4} (with \eqref{eq:ISAC-5} and
\eqref{eq:ISAC-6}) and \eqref{eq:ISAC-3} into \eqref{eq:ISAC-2}, the
mutual information\index{Mutual information} is expressed as
\begin{align}\label{eq:ISAC-7}
I(\pmb{X};\pmb{Y}|\pmb{H})=&T\log_2\left[\frac{\left(\pi e\right)^{N}\det(\pmb{H}\pmb{Q}_x\pmb{H}^H+\sigma^2\pmb{I}_N)}{\left(\pi e\right)^{N}\det(\sigma^2\pmb{I}_N)}\right]\nonumber\\
=&T\log_2\left[\det\left(\pmb{I}_N+\sigma^{-2}\pmb{H}\pmb{Q}_x\pmb{H}^H\right)\right]
\end{align}

Hence, the capability of the MIMO system\index{Capacity!MIMO system}
is
\begin{align}\label{eq:ISAC-8}
C(\pmb{H})=&\arg\max_{\pmb{x}_t:\textrm{Tr}(\pmb{Q}_x)\leq P_t}\left\{T^{-1}I(\pmb{X};\pmb{Y}|\pmb{H})=\log_2\left[\det\left(\pmb{I}_N+\sigma^{-2}\pmb{H}\pmb{Q}_x\pmb{H}^H\right)\right]\right\}
\end{align}
which has a unit of bits per symbol (transmission). In \eqref{eq:ISAC-8}, $\textrm{Tr}(\pmb{A})$ returns the trace of square matrix $\pmb{A}$ and $P_t$ is the total transmit power. The capacity of \eqref{eq:ISAC-8} can be achieved by designing $\pmb{Q}_x$ to match the channel $\pmb{H}$, as shown below.

Assume that the rank of $\pmb{H}$ is $G$. Then, applying the singular value decomposition (SVD)\index{singular value decomposition (SVD)} on $\pmb{H}$, yields 
\begin{align}\label{eq:ISAC-9}
\pmb{H}=\pmb{U}_h\pmb{\Sigma}_h^{1/2}\pmb{V}^H_h
\end{align}
where both $\pmb{U}_h\in\mathbb{C}^{N\times G}$ and $\pmb{V}_h\in\mathbb{C}^{M\times G}$ are sub-matrices of their unitary matrices $\pmb{U}$ and $\pmb{V}$, while the diagonal elements of $\pmb{\Sigma}_h=\textrm{diag}\left\{\lambda_1,\lambda_2,\ldots,\lambda_G\right\}$ are the non-zero eigenvalues of $\pmb{H}^H\pmb{H}$, where $\lambda_1\geq \lambda_2\geq \ldots\geq\lambda_G\geq 0$. Substituting \eqref{eq:ISAC-9} into  \eqref{eq:ISAC-7} gives
\begin{align}\label{eq:ISAC-10}
I(\pmb{X};\pmb{Y}|\pmb{H})=&T\log_2\left[\det\left(\pmb{I}_N+\sigma^{-2}\pmb{U}_h\pmb{\Sigma}_h^{1/2}\pmb{V}^H_h\pmb{Q}_x\pmb{V}_h\pmb{\Sigma}_h^{1/2}\pmb{U}_h^H\right)\right]\nonumber\\
=&T\log_2\left[\det\left(\pmb{I}_G+\sigma^{-2}\pmb{\Sigma}_h\pmb{V}^H_h\pmb{Q}_x\pmb{V}_h\right)\right]
\end{align}

The mutual information $I(\pmb{X};\pmb{Y}|\pmb{H})$ reaches maximum,
i.e., achieves the MIMO capacity, when
$\pmb{\beta}=\pmb{V}^H_h\pmb{Q}_x\pmb{V}_h$ is a diagonal matrix, with
the diagonal elements representing the power, obtained in
water-filling\index{Water-filling} principle, transmitted in the
corresponding virtual directions\index{Virtual direction} determined
by $\pmb{V}_h$~\cite{Lie-Liang-MC-CDMA-book}. Following the analysis
in Section 9.1 of \cite{Lie-Liang-MC-CDMA-book}, it can be shown that
\begin{align}\label{eq:ISAC-11}
\beta_g=P_t\left(\mu-\frac{\sigma^2}{\lambda_g}\right)^+,~g=1,2,\ldots,G
\end{align} 
where $(x)^+=x$ if $x> 0$, otherwise, $(x)^+=0$ if $x\leq 0$. In
\eqref{eq:ISAC-11}, the water level $\mu$ is obtained from the power
constraint of $\sum_{g=1}^G\beta_g\leq P_t$. Finally, upon
substituting the above results into \eqref{eq:ISAC-10}, the capacity
of the MIMO system\index{Capacity!MIMO system} with $M$ transmit and
$N$ receive antennas can be expressed as
\begin{align}\label{eq:ISAC-12}
C(\pmb{H})=\sum_{g=1}^G\log_2\left(\frac{P_t\mu\lambda_g}{\sigma^2}\right)^+
\end{align}
Explicitly, it encourages the virtual beams having higher gains to
convey more information. Note that for practical implementation, both
transmitter and receiver are required to employ the channel knowledge,
i.e., know $\pmb{H}$.

\index{MIMO!communications|)}
%%%%%%%
\subsection{MIMO Sensing}\label{subsection-6G-6.2.2}\index{MIMO!sensing|(}
%%%%%%%
%
\begin{figure}[tb]
  \begin{center}
    \includegraphics[width=0.65\linewidth]{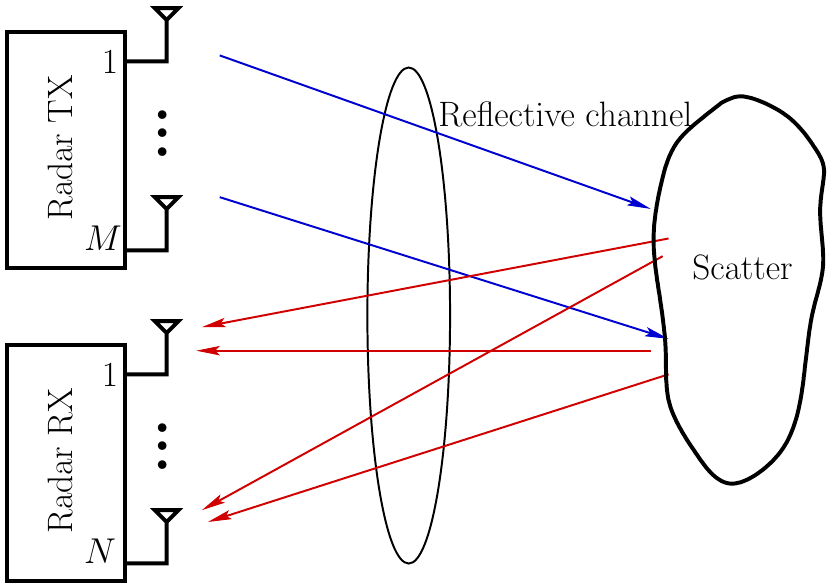}
 
  \end{center}
  \caption{Illustration of MIMO radar, where transmit and receive
    antennas may be co-located or separately located.}
\label{figure-MIMO-radar}
\end{figure}
Let us explicitly consider the MIMO radar\index{MIMO!radar} scheme as
shown in Fig.~\ref{figure-MIMO-radar}, but the model derived below is
general for many sensing systems. As shown in
Fig.~\ref{figure-MIMO-radar}, the radar system uses $M$ transmit
antennas to send probing signals\index{Probing signals}. After the
probing signals impinge on the scatters to be sensed, the echo signals
are received by $N$ receive antennas. Let $x_{tm}$, $m=1,\ldots,
M;~t=1,\ldots,T$, be a signal transmitted by the $m$th transmit
antenna at time $t$. Then, the corresponding signal observed by the
$n$th receive antenna can be written as
\begin{align}\label{eq:ISAC-13}
y_{tn}=\sum_{m=1}^Mx_{tm}h_{mn}+z_{tn},~n=1,2,\ldots,N;~t=1,2,\ldots,T
\end{align}
where $h_{mn}$ represents the channel experienced by the signal from the $m$ transmit antenna to the $n$th receive antenna, and $z_{tn}$ is complex Gaussian noise distributed with zero mean and a variance of $\sigma^2/2$ per dimension. Let $\pmb{x}_t=[x_{t1},x_{t2},\ldots,x_{tM}]^T$, $\pmb{h}_n=[h_{1n},h_{2n},\ldots,h_{Mn}]^T$. \eqref{eq:ISAC-13} can be written as
\begin{align}\label{eq:ISAC-14}
y_{tn}=\pmb{x}_{t}^T\pmb{h}_{n}+z_{tn},~n=1,2,\ldots,N;~t=1,2,\ldots,T
\end{align}
Now let $\pmb{y}_n=[y_{1n},y_{2n},\ldots,y_{Tn}]^T$ and
$\pmb{z}_n=[z_{1n},z_{2n},\ldots,z_{Tn}]^T$. Then, the signals
observed by the $n$th receive antenna over $T$ transmissions can be
expressed as
\begin{align}\label{eq:ISAC-15}
\pmb{y}_{n}=\pmb{X}\pmb{h}_{n}+\pmb{z}_{n},~n=1,2,\ldots,N
\end{align}
where $\pmb{X}=[\pmb{x}_1,\pmb{x}_2,\ldots,\pmb{x}_T]^T$, which is a
$(T\times M)$ matrix, i.e., $\pmb{X}\in\mathbb{C}^{T\times
  M}$. Finally, let $\pmb{Y}=[\pmb{y}_1,\pmb{y}_2,\ldots,\pmb{y}_N]$
and correspondingly,
$\pmb{Z}=[\pmb{z}_1,\pmb{z}_2,\ldots,\pmb{z}_N]$. The signals received
by the $N$ receive antennas over $T$ transmissions are collectively
expressed as
\begin{align}\label{eq:ISAC-16}
\pmb{Y}=\pmb{X}\pmb{H}+\pmb{Z}
\end{align}
where $\pmb{H}=[\pmb{h}_1,\pmb{h}_2,\ldots,\pmb{h}_N]\in\mathbb{C}^{M\times N}$.

Note that, when comparing \eqref{eq:ISAC-16} with \eqref{eq:ISAC-1}, we can observe that the representation of \eqref{eq:ISAC-16} for sensing has no difference from that of \eqref{eq:ISAC-1} for communication. It can be shown that \eqref{eq:ISAC-16} is simply the transpose of \eqref{eq:ISAC-1}. Hence, the analysis of MIMO sensing can be based on either \eqref{eq:ISAC-16} or \eqref{eq:ISAC-1}. Following analysis is based on \eqref{eq:ISAC-16}, which gives the representations that are slightly easy to follow.

Note furthermore that the representation of \eqref{eq:ISAC-16} is
general, many other sensing scenarios have similar representation. For
example, in the downlink sensing\index{Sensing!downlink} scenario,
where BS transmits sensing signals while a remote terminal receives
the scattered signals to sense the scatters, the observations obtained
at the remote terminal have the same representation of
\eqref{eq:ISAC-16}.

To carry out the following analysis, we assume that: a) $T\geq M\geq
N$ to make $\pmb{X}^T\pmb{X}^*$ full rank and that transmit
preprocessing is sufficient for optimum design; b) noise $\pmb{Z}$ is
not depended on transmission, the columns of which have a covariance
matrix $\sigma^2\pmb{I}_T$ and the rows of which have a covariance
matrix $\sigma^2\pmb{I}_N$; c) $\pmb{H}$ and $\pmb{Z}$ are
uncorrelated; d) the columns of $\pmb{H}$ obey the iid with zero mean
and a covariance matrix $\pmb{Q}_h\in\mathbb{C}^{M\times M}$, i.e.,
$E[\pmb{h}_n\pmb{h}_n^H]=\pmb{Q}_h$.

In \eqref{eq:ISAC-16}, $\pmb{Z}$ follows the complex Gaussian
distribution. Given $\pmb{X}$, $\pmb{Y}$ also follows the complex
Gaussian distribution, with the probability density function
(PDF)\index{Probability density function (PDF)} expressed
as~\cite{book-Steven-Kay-I,book:Narayan-Giri}
\begin{align}\label{eq:ISAC-17}
f(\pmb{Y}|\pmb{X})=&\prod_{n=1}^Nf(\pmb{y}_n|\pmb{X})=\prod_{n=1}^N\frac{1}{(\pi\sigma^2)^T}\exp\left(-\pmb{y}^H_n\pmb{R}_{\pmb{y}_n}^{-1}\pmb{y}_n\right)
\end{align}
where $\pmb{R}_{\pmb{y}_n}\in\mathbb{C}^{T\times T}$ is the covariance
matrix of $\pmb{y}_n$ conditioned on a given $\pmb{X}$, having an
expression of
\begin{align}\label{eq:ISAC-18}
\pmb{R}_{\pmb{y}_n}=E\left[\pmb{y}_n\pmb{y}_n^H\right]=\pmb{X}\pmb{Q}_h\pmb{X}^H+\sigma^2\pmb{I}_T
\end{align}

Note that $\pmb{Z}$'s PDF is give by \eqref{eq:ISAC-17} with $\pmb{R}_{\pmb{y}_n}$ replaced by $\sigma^2\pmb{I}_T$.

For optimal sensing design, given the total transmit power and the
second order statistics of $\pmb{Q}_h$, the objective is to design the
$\pmb{X}$ to satisfy the certain characteristics (to be determined
below), so that the mutual information\index{Mutual information}
between $\pmb{H}$, which embeds the sensing information, and the
received observation $\pmb{Y}$ is maximum. When $\pmb{X}$ is designed
in such a way, the receiver is capable of deriving the maximal
information about $\pmb{H}$ and hence, achieving the best sensing
performance. The mutual information can be referred to as the {\em
  estimation rate}~\cite{7279172,9737357}, which has a unit, such as,
bits per transmission, explaining the amount of information about the
channel attained by receiver from one transmission.

According to \cite{book:Thomas-Cover-Information-Theory}, the mutual information
between $\pmb{H}$ and $\pmb{Y}$ conditioned on $\pmb{X}$ can be
formulated as
\begin{align}\label{eq:ISAC-19}
I(\pmb{H};\pmb{Y}|\pmb{X})=&h(\pmb{Y}|\pmb{X})-h(\pmb{Y}|\pmb{X},\pmb{H})\nonumber\\
=&h(\pmb{Y}|\pmb{X})-h(\pmb{Z})
\end{align}
In \eqref{eq:ISAC-19},
\begin{align}\label{eq:ISAC-20}
h(\pmb{Z})=&-\int f(\pmb{Z})\log_2f(\pmb{Z})d\pmb{Z}\nonumber\\
=&N\log_2\left[(\pi e)^T\det(\sigma^2\pmb{I}_T)\right]
\end{align}
\begin{align}\label{eq:ISAC-21}
h(\pmb{Y}|\pmb{X})=&-\int f(\pmb{Y}|\pmb{X})\log_2f(\pmb{Y}|\pmb{X})d\pmb{Y}\nonumber\\
=&N\log_2\left[(\pi e)^T\det(\pmb{R}_{\pmb{y}_n})\right]\nonumber\\
=&N\log_2\left[(\pi e)^T\det\left(\pmb{X}\pmb{Q}_h\pmb{X}^H+\sigma^2\pmb{I}_T\right)\right]
\end{align}
Hence, substituting \eqref{eq:ISAC-20} and \eqref{eq:ISAC-21} into \eqref{eq:ISAC-19} yields 
\begin{align}\label{eq:ISAC-22}
I(\pmb{H};\pmb{Y}|\pmb{X})=&N\log_2\left[\frac{\left(\pi e\right)^{T}\det(\pmb{X}\pmb{Q}_h\pmb{X}^H+\sigma^2\pmb{I}_T)}{\left(\pi e\right)^{T}\det(\sigma^2\pmb{I}_T)}\right]\nonumber\\
=&N\log_2\left[\det\left(\pmb{I}_T+\sigma^{-2}\pmb{X}\pmb{Q}_h\pmb{X}^H\right)\right]\\
\label{eq:ISAC-23}
=&N\log_2\left[\det\left(\pmb{I}_M+\sigma^{-2}\pmb{Q}_h\pmb{X}^H\pmb{X}\right)\right]
\end{align}
where from \eqref{eq:ISAC-22} to \eqref{eq:ISAC-23}, the property of
$\det(\pmb{I}+\pmb{AB})=\det(\pmb{I}+\pmb{BA})$ is applied. Hence, the
maximal estimation rate, or the sensing capacity, is obtained by
solving the optimization problem of
\begin{align}\label{eq:ISAC-24}
C(\pmb{Q}_h)=\arg\max_{\frac{1}{T}\textrm{Tr}\left(\pmb{X}\pmb{X}^H\right)\leq P_t}\left\{\frac{1}{T}I(\pmb{H};\pmb{Y}|\pmb{X})=\frac{N}{T}\log_2\left[\det\left(\pmb{I}_M+\sigma^{-2}\pmb{Q}_h\pmb{X}^H\pmb{X}\right)\right]\right\}
\end{align}
which has the unit of bits per transmission. 

To maximize the estimation rate in \eqref{eq:ISAC-24}, the sensing signals sent from transmit antennas and corresponding transmit power can be derived from solving the optimization problem~\cite{5467189}
\begin{align}\label{eq:ISAC-25}
\tilde{\pmb{X}}=&\arg\max_{\pmb{X}}\left\{\log_2\left[\det\left(\pmb{I}_M+\sigma^{-2}\pmb{Q}_h\pmb{X}^H\pmb{X}\right)\right]\right\}\nonumber\\
 s.t.& ~~\textrm{Tr}\left(\pmb{X}\pmb{X}^H\right)\leq TP_t
\end{align}
where {\em s.t.} is for `subject to'. Assume that the rank of $\pmb{Q}_h$ is $G$ and the $G$ eigenvalues are $\lambda_1\geq \lambda_2\geq\ldots\geq\lambda_G>0$. Then, $\pmb{Q}_h$ can be decomposed as
\begin{align}\label{eq:ISAC-26}
\pmb{Q}_h=\pmb{V}_h\pmb{\Sigma}_h\pmb{V}_h^H
\end{align} 
where the columns of $\pmb{V}_h\in\mathbb{C}^{M\times G}$ are the $G$
eigenvectors\index{Eigenvectors} of $\pmb{Q}_h$ corresponding to its
$G$ non-zero eigenvalues\index{Eigenvalues},
$\pmb{\Sigma}_h=\textrm{diag}\{\lambda_1,\lambda_2,\ldots,\lambda_G\}$. Substituting
\eqref{eq:ISAC-26} into \eqref{eq:ISAC-26} gives
\begin{align}\label{eq:ISAC-27}
\tilde{\pmb{X}}=&\arg\max_{\pmb{X}}\left\{\log_2\left[\det\left(\pmb{I}_M+\sigma^{-2}\pmb{V}_h\pmb{\Sigma}_h\pmb{V}_h^H\pmb{X}^H\pmb{X}\right)\right]\right\}\nonumber\\
=&\arg\max_{\pmb{X}}\left\{\log_2\left[\det\left(\pmb{I}_G+\sigma^{-2}\pmb{\Sigma}_h(\pmb{X}\pmb{V}_h)^H\pmb{X}\pmb{V}_h\right)\right]\right\}\nonumber\\
 s.t.&~~ \textrm{Tr}\left((\pmb{X}\pmb{V}_h)^H\pmb{X}\pmb{V}_h\right)\leq TP_t
\end{align}
Hence, to maximize the mutual information, $\pmb{\beta}=(\pmb{X}\pmb{V}_h)^H\pmb{X}\pmb{V}_h$ should be a $(G\times G)$ diagonal matrix. Explicitly, this can be achieved by designing $\pmb{X}$ to satisfy
\begin{align}\label{eq:ISAC-28}
\pmb{X}=\pmb{U}_x\pmb{\beta}^{1/2}\pmb{V}_h^H
\end{align}
where $\pmb{U}_x\in\mathbb{C}^{T\times G}$ is constituted by $G$ columns of unitary matrix $\pmb{U}\in\mathbb{C}^{T\times T}$, and $\pmb{\beta}=\textrm{diag}\{\beta_1,\beta_2,\ldots,\beta_G\}$ accounts for the transmit power. 

From \eqref{eq:ISAC-28} we can be inferred that, physically, to
achieve the best sensing performance, the signals transmitted by
different transmit antennas over $T$ transmissions should be
orthogonal to each other, while during each transmission, the sensing
signals should be sent in the directions determined by the
eigenvectors of the channel, i.e., $\pmb{Q}_h$, associated with
appropriately allocated power. Assume that $\pmb{X}$ is designed in
such a way. Then, what left is to find the optimal $\pmb{\beta}$,
which can be obtained by solving the problem
\begin{align}\label{eq:ISAC-29}
\tilde{\pmb{\beta}}=&\arg\max_{\pmb{\beta}}\left\{\log_2\left[\det\left(\pmb{I}_G+\sigma^{-2}\pmb{\Sigma}_h\pmb{\beta}\right)\right]\right\}\nonumber\\
 s.t.&~~ \sum_{g=1}^G\beta_g\leq TP_t
\end{align}
This is a typical water-filling
power-allocation\index{Power-allocation!water-filling} problem, whose
solutions can be found in references, such as in
\cite{MIMO-Telatar-I,Lie-Liang-MC-CDMA-book}, which are
\begin{align}\label{eq:ISAC-30}
\beta_g=TP_t\left(\mu-\frac{\sigma^2}{\lambda_g}\right)^+,~g=1,2,\ldots,G
\end{align} 
where the water level $\mu$ is obtained from the power constraint of
$\sum_{g=1}^G\beta_g\leq TP_t$. When substituting the above results
into \eqref{eq:ISAC-29}, the maximum sensing rate, i.e., sensing
capacity, achieved by the MIMO sensing system with $M$ transmit and
$N$ receive antennas and using $T$ transmissions can be expressed as
\begin{align}\label{eq:ISAC-31}
C(\pmb{Q}_h)=\frac{N}{T}\sum_{g=1}^G\log_2\left(\frac{TP_t\mu\lambda_g}{\sigma^2}\right)^+
\end{align}

Note that the estimation rate of \eqref{eq:ISAC-31} is the rate obtained from the average over $T$ transmissions. $P_t$ has the meaning of energy per transmission. Hence, when given that the channel $\pmb{H}$ does not change during sensing, it is sensible that more transmissions result in higher reliability of the sensing for $\pmb{H}$. \eqref{eq:ISAC-31} also shows that the sensing reliability of $\pmb{H}$, measured by estimation rate, linearly increases with the number of receive antennas $N$. The impact of the number of transmit antennas on the sensing reliability is not explicit. However, when $T>M$, increasing $M$ in general results in the increase of $G$, which also results in the increase of estimation rate and hence, enhances the reliability of channel sending.      

\index{MIMO!sensing|)}
%%%
\section{Resource Optimization in MIMO ISAC Systems}\label{subsection-6G-6.3}\index{MIMO ISAC!resource optimization|(}
%%%
%
\begin{figure}[tb]
  \begin{center}
    \includegraphics[width=0.65\linewidth]{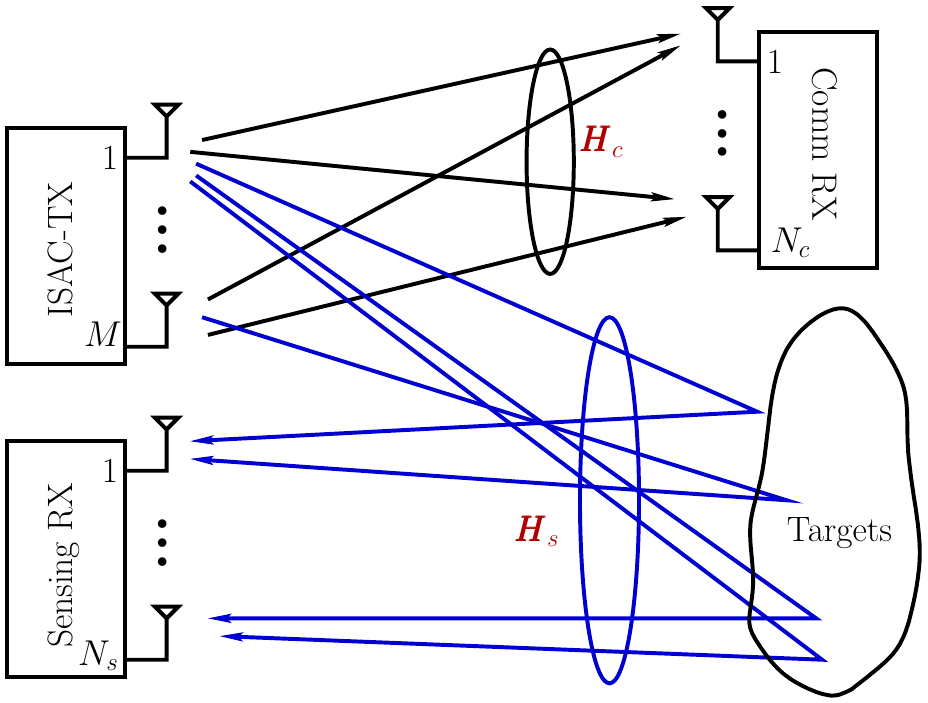}
 
  \end{center}
  \caption{Illustration of a MIMO ISAC system, where the sensing receive
    antennas may be co-located with the ISAC transmitter or communication receiver, or separately located.}
\label{figure-MIMO-ISAC}
\end{figure}

Having analyzed the signal design for communications and sensing,
respectively, in Sections \ref{subsection-6G-6.2.1} and
\ref{subsection-6G-6.2.2}, now we consider the signal design for joint
communication and sensing in MIMO ISAC systems. For this purpose, an
MIMO ISAC system with a diagram as shown in
Fig.~\ref{figure-MIMO-ISAC} is considered. In this system, the ISAC
transmitter uses $M$ antennas to transmit ISAC signals, the
communication receiver employs $N_c$ antennas for information
detection, while the sensing receiver utilizes $N_s$ antennas for
targets (channel) sensing. The channel matrix from ISAC transmitter to
communication receiver is expressed as
$\pmb{H}_c\in\mathbb{C}^{N_c\times M}$, and the channel matrix from
ISAC transmitter to sensing receiver is expressed as
$\pmb{H}_s\in\mathbb{C}^{M\times N_s}$.

Note that this model can be readily modified for different
scenarios. First, the sensing receiver may be co-located with the ISAC
transmitter, with the communication receiver, or located away from
both the ISAC transmitter and communication receiver. If the sensing
receiver is also the communication receiver, we have
$\pmb{H}_s=\pmb{H}_c^T$. Otherwise, $\pmb{H}_s\neq\pmb{H}_c^T$. If the
sensing targets are the same of the communication receivers, the
information, such as locations of communication receivers, obtained
from sensing can be exploited for the transceiver optimization of
communication. Second, the model may be generalized to consider
multiple communication receivers, in the form of multiuser MIMO
(MU-MIMO)\index{MIMO!multiuser}. In this scenario, BS acts as the ISAC
transmitter, which sends information to a number of downlink users,
while simultaneously sends signals for sensing some targets that may
or may not be the downlink users. In this case, again, the sensing
receiver may be the BS itself or another terminal dedicated for
sending.

Assume that a frame consisting of $T$ transmissions is considered. Following Sections \ref{subsection-6G-6.2.1} and \ref{subsection-6G-6.2.2}, the observations received by communication receiver can be represented as  
\begin{align}\label{eq:ISAC-32}
\pmb{Y}_c=\pmb{H}_c\pmb{X}+\pmb{N}
\end{align}
where $\pmb{Y}_c\in\mathbb{C}^{N_c\times T}$, $\pmb{X}\in\mathbb{C}^{M\times T}$ and noise $\pmb{N}\in\mathbb{C}^{N_c\times T}$. The observations received by sensing receiver can be written as 
\begin{align}\label{eq:ISAC-33}
\pmb{Y}_s=\pmb{X}^T\pmb{H}_s+\pmb{Z}
\end{align}
where $\pmb{Y}_s\in\mathbb{C}^{T\times N_s}$ and noise $\pmb{Z}\in\mathbb{C}^{T\times N_s}$.

Resource optimization in the above MIMO ISAC systems is required to
design a $\pmb{X}$ in \eqref{eq:ISAC-32} and \eqref{eq:ISAC-33}, which
allows both communication and sensing to attain the highest possible
efficiency. From Sections \ref{subsection-6G-6.2.1} and
\ref{subsection-6G-6.2.2} we are inferred that this includes designing
the virtual beams and allocating the corresponding power to these
virtual beams. Separately doing these, both communication and sensing
are capable of achieving their optimum design to maximize mutual
information\index{Mutual information}. However, the analyses in
Sections \ref{subsection-6G-6.2.1} and \ref{subsection-6G-6.2.2} show
that the optimum signals for communications and channel sensing have
different requirements. Specifically, as shown in
Section~\ref{subsection-6G-6.2.1}, the communication signals should be
designed to make $\pmb{\beta}=\pmb{V}^H_h\pmb{Q}_x\pmb{V}_h$ a
diagonal matrix with the diagonal elements of $\pmb{\beta}$ determined
in the principle of water-filling. This can be fulfilled by designing
the transmit signals to have an autocorrelation matrix
$\pmb{Q}_x=E[\pmb{x}_t\pmb{x}_t^H]$ that can be decomposed into
$\pmb{Q}_x=\pmb{V}_h\pmb{\beta}\pmb{V}^H_h$. It means that the
transmit signals should be sent in the directions determined by the
eigen-beams\index{Eigen-beams} determined by the propagation channel
$\pmb{H}_c$, and more power is assigned to a better eigen-beam having
a higher beam-gain, enabling it to convey more information.

By contrast, as shown in Section~\ref{subsection-6G-6.2.2}, the
sensing signals should be designed to make
$\pmb{\beta}=(\pmb{X}^T\pmb{V}_h)^H\pmb{X}^T\pmb{V}_h$ a diagonal
matrix\footnote{Following \eqref{eq:ISAC-33}, $\pmb{X}$ in
  Section~\ref{subsection-6G-6.2.2} is replaced by $\pmb{X}^T$.}, with
the diagonal elements of $\pmb{\beta}$ determined in the principle of
water-filling. As shown in \eqref{eq:ISAC-28}, $\pmb{X}^T$ can be
designed to satisfy $\pmb{X}^T=\pmb{U}_x\pmb{\beta}^{1/2}\pmb{V}_h^H$,
where $\pmb{X}^T\in\mathbb{C}^{T\times M}$ is constructed by the
sensing signals sent from $M$ transmit antennas over $T$ transmission
intervals, the $m$th column of $\pmb{X}^T$ is the signals sent by the
$m$th transmit antenna with $T$ transmissions. Therefore, to achieve
the optimum sensing of channel $\pmb{H}_s$, first, signals sent from
different transmit antennas over $T$ transmission intervals should be
orthogonal to each other, to obtain $\pmb{U}_x$ seen in
$\pmb{X}^T=\pmb{U}_x\pmb{\beta}^{1/2}\pmb{V}_h^H$. Second, within each
transmission interval, signals send from $M$ transmit antennas should
be in the directions determined by the eigenvectors of
$\pmb{Q}_h=E[\pmb{h}_n\pmb{h}_n^H]$, reflected by $\pmb{V}_h$ in
$\pmb{X}^T=\pmb{U}_x\pmb{\beta}^{1/2}\pmb{V}_h^H$. Third, more sensing
power should be assigned to transmit signals in an
eigen-direction\index{Eigen-direction} that is capable of providing
more information about the channel $\pmb{H}_s$ to receiver, which is
explained by $\pmb{\beta}$, obtained in water-filling principle, in
$\pmb{X}^T=\pmb{U}_x\pmb{\beta}^{1/2}\pmb{V}_h^H$.

Therefore, the optimum signals designed for communication cannot also
be optimum for channel sensing. Similarly, the optimum signals
designed for channel sensing cannot simultaneously be optimum for
information delivery. Hence, in the design of ISAC systems, a
trade-off between sensing and communication performance has to be
made. With this regard, the objective function for a general
optimization problem in ISAC may be formulated as the weighted sum of
the mutual information of both communication and
sensing~\cite{9303435,7970102,9540344}, expressed as
\begin{align}\label{eq:ISAC-34}
J(\pmb{X})=\frac{\rho}{C(\pmb{H}_c)}I(\pmb{X};\pmb{Y}|\pmb{H}_c)+\frac{1-\rho}{C(\pmb{Q}_h)}I(\pmb{H}_s;\pmb{Y}|\pmb{X})
\end{align}
where $\pmb{H}_c$ represents the communication channel and $\pmb{H}_s$
is the channel being sensed, which may be or may not be the same,
depended on where the sensing is operated~\cite{9540344}, as
previously mentioned. In \eqref{eq:ISAC-34}, $C(\pmb{H}_c)$ and
$C(\pmb{Q}_h)$ are the maximum rates obtained in
Sections~\ref{subsection-6G-6.2.1} and \ref{subsection-6G-6.2.2} by
considering only communication and sensing, respectively, and $\rho$
is the weight coefficient.

The function of \eqref{eq:ISAC-34} is the sum of two concave
functions. Hence, it is concave, which can be solved under the
constraint of total transmit power, for example, with the aid of the
Karush-Kuhn-Tucker (KKT)\index{Karush-Kuhn-Tucker (KKT) conditions}
conditions~\cite{9303435}. It is shown that the solutions are in the
principle of water-filling, which jointly address the effects of
$\pmb{H}_c$ and $\pmb{Q}_x$ in communication, as well as the effect of
$\pmb{Q}_h$ and the requirements of $\pmb{X}$ in sensing.

To optimize \eqref{eq:ISAC-34}, the covariance matrix $\pmb{Q}_h$ of
the sensing channel is assumed to be {\em a-priori} knowledge, which
may not be the case in most application scenarios, unless static
transceivers and environments are considered. In practice,
communication/sensing terminals may be in move and environments are
typically time-varying. Furthermore, often only a few of targets are
sensed. Hence, the sensing channel can hardly be modeled as a
stationary process that gives a sufficiently stable covariance matrix
$\pmb{Q}_h$ for its estimation and exploitation. In this case,
$\pmb{X}$ needs to be designed without assuming $\pmb{Q}_h$. There are
various methods for implementation, below are some extended from
\cite{8386661}.

Assume that for communication, the data symbols sent to communication
receiver are expressed as $\pmb{C}\in\mathbb{C}^{K\times T}$, where
$K$ is the number of data streams or the number of downlink users. In
this case, $\pmb{X}_c$ for communication can be written as
$\pmb{X}_c=\pmb{F}\pmb{C}$, with $\pmb{F}\in\mathbb{C}^{M\times K}$ is
the transmitter preprocessing matrix.\index{Transmitter preprocessing}
Then, we can modify \eqref{eq:ISAC-32} to the form of
\begin{align}\label{eq:ISAC-35}
\pmb{Y}_c=\pmb{C}+(\pmb{H}_c\pmb{X}_c-\pmb{C})+\pmb{N}
\end{align}
which explicitly shows the interference of $(\pmb{H}_c\pmb{X}_c-\pmb{C})$. Hence, maximizing the performance of communication needs to minimize the interference power of
\begin{align}\label{eq:ISAC-36}
P_I(\pmb{X}_c)=\|\pmb{H}_c\pmb{X}_c-\pmb{C}\|^2_{F}
\end{align}
where $\|\pmb{A}\|_F$ denotes the Frobenius norm of $\pmb{A}$. 

Now, assume that only sensing is considered, which yields a desired $\pmb{X}_s\in\mathbb{C}^{M\times T}$ having the autocorrelation matrix 
\begin{align}\label{eq:ISAC-37}
\pmb{R}_s=\pmb{X}_s\pmb{X}_s^H
\end{align}
Assume that the total transmit power is $TP_t$, meaning that each transmission is constrained by a total power of $P_t$. Then, when the joint design of communication and sensing is considered, the objective will be to design the $\pmb{X}_c$ (or $\pmb{F}$) such that $P_I(\pmb{X}_c)$ is minimum, mathematically described as
\begin{subequations}\label{eq:ISAC-38}
\begin{align}
\tilde{\pmb{X}}_c=&\arg\min_{\pmb{X}_c}\left\{\|\pmb{H}_c\pmb{X}_c-\pmb{C}\|^2_{F}\right\}\\
s.t. &~~\pmb{X}_c\pmb{X}_c^H=\pmb{R}_s
\end{align}
\end{subequations}
Note that, Problem \eqref{eq:ISAC-38} is non-convex due to the equality constraint, but it can be solved with the aid of SVD~\cite{8386661}\index{Singular value decomposition (SVD)}.

The problem \eqref{eq:ISAC-38} imposes a strict constraint of
\eqref{eq:ISAC-38}(b) on the sensing waveforms. In this case, the
derived $\tilde{\pmb{X}}_c$ may result in a big penalty of
communication performance. To strike a promising trade-off between the
performance of sensing and that of communication, the joint design may
consider a trade-off between the minimization of interference and the
ideality of sensing waveform, with the optimization problem described
as
\begin{align}\label{eq:ISAC-39}
\tilde{\pmb{X}}_c=&\arg\min_{\pmb{X}_c}\left\{\rho\|\pmb{H}_c\pmb{X}_c-\pmb{C}\|^2_{F}+(1-\rho)\|\pmb{X}_c-\pmb{X}_s\|^2_F\right\}\nonumber\\
s.t. &~~\textrm{Tr}\left(\pmb{X}_c\pmb{X}_c^H\right)=TP_t
\end{align}
where $0\leq \rho\leq 1$ is the weight coefficient; $\rho=0$
implements optimum sensing without caring about communication,
$\rho=1$ achieves optimum communication regardless of the performance
of sensing, while $0< \rho< 1$ attains a trade-off between them. The
optimization of \eqref{eq:ISAC-39} is a Pareto
optimization\index{Optimization!Pareto}, which can be transformed to a
semidefinite programming (SP)\index{Optimization!Semidefinite
  programming (SP) problem} problem, solved by the semidefinite
relaxation (SDR)\index{Semidefinite relaxation
  (SDR)}~\cite{8386661,Lie-Liang-MC-CDMA-book}.

The above optimization problems in \eqref{eq:ISAC-38} and
\eqref{eq:ISAC-39} assume the constraint of total transmit
power. Instead, when the constraint per antenna transmit power is
assumed, \eqref{eq:ISAC-39} can alternatively be stated as
\begin{align}\label{eq:ISAC-40}
\tilde{\pmb{X}}_c=&\arg\min_{\pmb{X}_c}\left\{\rho\|\pmb{H}_c\pmb{X}_c-\pmb{C}\|^2_{F}+(1-\rho)\|\pmb{X}_c-\pmb{X}_s\|^2_F\right\}\nonumber\\
s.t. &~~\textrm{diag}^{-1}\left(\pmb{X}_c\pmb{X}_c^H\right)=\frac{TP_t}{M}\pmb{1}_M
\end{align}
where $\textrm{diag}^{-1}\left(\pmb{A}\right)$ denotes a vector formed by
the diagonal elements of $\pmb{A}$, and $\pmb{1}_M$ is a $M$-length
vector with unity elements.

Additionally, as radar signals desire the constant modulus
waveforms\index{Constant modulus waveform}, further constraint
accounting for this can be added, having the optimization problem
described as
\begin{align}\label{eq:ISAC-41}
\tilde{\pmb{X}}_c=&\arg\min_{\pmb{X}_c}\left\{\rho\|\pmb{H}_c\pmb{X}_c-\pmb{C}\|^2_{F}+(1-\rho)\|\pmb{X}_c-\pmb{X}_s\|^2_F\right\}\nonumber\\
s.t. &~~|\pmb{X}_c(i,j)|=\sqrt{\frac{P_t}{M}},~~\forall i,j
\end{align}
which can be solved by various methods, when different kinds of norms are considered. 

\index{MIMO ISAC!resource optimization|)}
%%%
\section{ISAC mmWave Systems}\label{subsection-6G-6.4}\index{ISAC!mmWave|(}
%%%

The MIMO ISAC model analyzed in Section~\ref{subsection-6G-6.3} is
general and can be applied with most MIMO systems either directly or
after slight modification. In this section, the ISAC in conjunction
with millimeter wave (mmWave) communications is specifically treated.

Owing to their large bandwidth, millimeter wavelength and its resulted
small profile massive antenna arrays, mmWave systems have the
potential to provide very high data rates for communications. On the
other side, mmWave signals provide high angular resolution and a
mmWave channel usually consists of only a few of multipath components,
which are hence beneficial for environment sensing and target
positioning. Owing to the above-mentioned, ISAC in mmWave systems has
been regarded as one of the important research topics towards the
future 6G wireless
systems~\cite{9737357,8999605,9540344,8550811,9131843,8288677,9898900,10275023,10316594}

As in Section~\ref{subsection-6G-6.3}, let us consider a downlink
mmWave system\index{mmWave!downlink}, where BS employs an array of $M$
elements, a sensing receiver has an array of $N_s$ elements and $K\geq
1$ communications users are supported, with each having an array of
$N_k$ elements. Note that in practical mmWave systems, $M$ is typical
significantly larger than the total number of data steams of $K$
users, forming the so-called massive MIMO\index{Massive MIMO}, which
is also assumed in the following analysis. Then, following
\eqref{eq:ISAC-32} and \eqref{eq:ISAC-33}, the observation signals
derived at a communication user and at the sensing receiver have the
forms of
\begin{align}\label{eq:ISAC-42}
\pmb{y}_{k,t}=&\pmb{H}_{k,t}\pmb{x}_t+\pmb{n}_{k,t},~k=1,2,\ldots,K\\
\label{eq:ISAC-43}
\pmb{y}^T_{s,t}=&\pmb{x}^T_t\pmb{H}_{s,t}+\pmb{z}^T_{s,t}
\end{align}
for $t=1,2,\ldots,T$, where the symbol-level observations are
represented. In \eqref{eq:ISAC-42} and \eqref{eq:ISAC-43},
$\pmb{y}_{k,t}\in\mathbb{C}^{N_k\times 1}$ and
$\pmb{y}_{s,t}\in\mathbb{C}^{N_s\times 1}$ are respectively the
observations obtained at a communication user and the sensing
receiver. $\pmb{n}_{k,t}\in\mathbb{C}^{N_k\times 1}$ and
$\pmb{z}_{s,t}\in\mathbb{C}^{N_s\times 1}$ account for the background
noise and possibly other interference, such as clutter interference in
sensing. $\pmb{H}_{k,t}\in\mathbb{C}^{N_k\times M}$ and
$\pmb{H}_{s,t}\in\mathbb{C}^{M\times N_s}$, and
$\pmb{H}_{s,t}^T\equiv\pmb{H}_{k,t}$, meaning that $\pmb{H}_{s,t}^T$
is equivalent to $\pmb{H}_{k,t}$ in terms of their
construction. Finally, $\pmb{x}_t\in\mathbb{C}^{M\times 1}$ contains
both the information to $K$ users and the probing signals for channel
environment sensing.

Note that, as previously mentioned associated with \eqref{eq:ISAC-32} and \eqref{eq:ISAC-33}, \eqref{eq:ISAC-42} and \eqref{eq:ISAC-43} are equivalent in terms of their structures, i.e., $\pmb{y}_{s,t}\equiv \pmb{y}_{k,t}$. \eqref{eq:ISAC-43} (also \eqref{eq:ISAC-33}) might be slightly preferred for sensing processing, as, first, $\pmb{H}_{s,t}$ ($\pmb{H}_{s}$) is the variable to be estimated and, second, it is desirable that the probing signals sent from different antenna elements are orthogonal over $T$ transmission intervals. Otherwise, \eqref{eq:ISAC-42} and \eqref{eq:ISAC-43} (or \eqref{eq:ISAC-32} and \eqref{eq:ISAC-33}) have no difference. Due to their equivalence, \eqref{eq:ISAC-42} will be focused on in the following analysis. 

While \eqref{eq:ISAC-42} and \eqref{eq:ISAC-43} have the structures of general MIMO communication and MIMO sensing, respectively, the matrices/vectors invoked have their specific characteristics, when mmWave signaling is considered.

First, in mmWave systems, beamforming\index{Beamforming} is typically
carried out at both transmitter and receiver, referred to as
precoding\index{Precoding!transmitter} at transmitter and
combining\index{Combining!receiver} at receiver. Correspondingly,
\eqref{eq:ISAC-42} can be represented as
\begin{align}\label{eq:ISAC-44}
\pmb{y}_{k,t}=&\pmb{W}_{k,t}^H\pmb{H}_{k,t}\pmb{F}_t\pmb{s}_t+\pmb{n}_{k,t},~k=1,2,\ldots,K;~t=1,2,\ldots,T
\end{align}
where the same matrices $\pmb{y}_{k,t}$ and $\pmb{n}_{k,t}$ are used
for simplicity, but they may have the dimension and property
differences. In \eqref{eq:ISAC-44}, $\pmb{s}_t$ contains the
information to users and possibly extra probing signals, as seen below
in the special cases. At moment, we assume
$\pmb{s}_t\in\mathbb{C}^{C\times 1}$. Then,
$\pmb{F}_t\in\mathbb{C}^{M\times C}$ is the precoding
matrix\index{Precoding!matrix}, which can be implemented purely in
digital domain or in hybrid analog-digital domain. In the later case,
usually, a near-optimum $\pmb{F}_t^o$ is first designed in digital
domain. Then, the digital baseband precoder, $\pmb{F}_{t,BB}^o$, and
radio-frequency (RF) analogue precoder, $\pmb{F}_{t,RF}^o$, are
designed as
\begin{align}\label{eq:ISAC-45}
\{\pmb{F}_{t,RF}^o,\pmb{F}_{t,BB}^o\}=\arg\min_{\pmb{F}_{t,RF},\pmb{F}_{t,BB}}\left\{\|\pmb{F}_t^o-\pmb{F}_{t,RF}\pmb{F}_{t,BB}\|_F^2\right\}
\end{align}
under the certain constraints depended on the design, where the
dimensions of $\pmb{F}_{t,RF}^o$ and $\pmb{F}_{t,BB}^o$ are depended
on the number of data streams and the number of RF
chains~\cite{7400949}. Similarly, assume that $I_{k}$ data streams
are transmitted to user $k$. Then,
$\pmb{W}_{k,t}\in\mathbb{C}^{N_k\times I_k}$ is the combiner of user
$k$, which can also be implemented purely in digital domain or in
hybrid analog-digital domain. Furthermore, when implemented in hybrid
analog-digital domain, $\pmb{W}_{k,t,RF}^o$ and $\pmb{W}_{k,t,BB}^o$
can be designed similarly as the precoders in \eqref{eq:ISAC-45}, with
their dimensions determined by the number of data streams and that of
RF chains of user $k$.

Second, the mmWave channel has a range of characteristics that are
beneficial for ISAC implementation. Assume that there are at most $L$
physical propagation paths\footnote{If the number of paths from BS to
  user $k$ is less than $L$, some paths can be simply set to have a
  gain of zero.} from BS to user $k$. For simplicity, we assume
narrow-band transmission and that Doppler-shift is mild, making the
mmWave channel neither time-selective nor frequency-selective. Then,
the channel matrix $\pmb{H}_{k,t}$ can be represented
as~\cite{8827589,7400949}
\begin{align}\label{eq:ISAC-46}
\pmb{H}_{k,t}=&\sum_{l=1}^L\alpha_l^{(k)}e^{-j2\pi f\tau_l^{(k)}}e^{j2\pi tf_{D,l}^{(k)}T_s}\pmb{a}(N_k,\phi_{k,l})\pmb{a}^T(M,\theta_{BS,l}),\nonumber\\
&~~t=1,2,\ldots,T
\end{align}
For the $l$th propagation path, the terms in \eqref{eq:ISAC-46} have
the meaning as follows: $\alpha_l^{(k)}$ is the propagation path gain,
$f$ is the carrier frequency, $\tau_l^{(k)}$ is the propagation delay
from BS to user $k$, $f_{D,l}^{(k)}$ is the Doppler
frequency-shift\index{Doppler frequency-shift}, which is a function of
the relative moving between BS and user $k$, $T_s$ is the symbol
duration (or sampling interval), $\theta_{BS,l}$ is the
angle-of-departure (AoD)\index{Angle-of-depature (AoD)} at BS,
$\phi_{k,l}$ is the angle-of-arrival (AoA)\index{Angle-of-arrival
  (AoA)} at user $k$, $\pmb{a}^T(M,\theta_{BS,l})$ is the response
vector of the transmit array at BS, and $\pmb{a}(N_k,\phi_{k,l})$ is
the response vector of the receive array at user $k$. Usually, we
assume that these terms keep invariant over $T$ transmission intervals
of one sensing period.

For example, when assuming a uniform linear array (ULA)\index{Uniform
  linear array (ULA)} having $N$ elements, where $N$ is either $N_k$
or $M$, the array response vector (or so-called steering vector,
beamforming vector, etc) can be represented as~\cite{7400949}
\begin{align}\label{eq:ISAC-47}
\pmb{a}(N,\theta)=\left[1,e^{-j2\pi\vartheta},e^{-j4\pi\vartheta},\ldots,e^{-j2(N-1)\pi\vartheta}\right]^T/\sqrt{N}
\end{align}
where the normalized $\vartheta$ is related to the physical angle $\theta\in[-\pi/2,\pi/2]$ as
\begin{align}\label{eq:ISAC-48}
\vartheta=\frac{d\sin(\theta)}{\lambda}
\end{align}
where $\lambda$ is wavelength and $d$ is antenna spacing. 

For convenience, \eqref{eq:ISAC-46} can be written as
\begin{align}\label{eq:ISAC-49}
\pmb{H}_{k,t}=&\pmb{A}_k\pmb{\Sigma}_{k,t}\pmb{A}_{BS}^T\nonumber\\
=&\pmb{A}_k\pmb{\alpha}_k\pmb{D}_k\pmb{O}_{k,t}\pmb{A}_{BS}^T
\end{align} 
where $\pmb{\Sigma}_{k,t}=\pmb{\alpha}_k\pmb{D}_k\pmb{O}_{k,t}$ and by definition,
\begin{align}\label{eq:ISAC-50}
\pmb{A}_k=&\left[\pmb{a}(N_k,\phi_{k,1}),\pmb{a}(N_k,\phi_{k,2}),\ldots,\pmb{a}(N_k,\phi_{k,L})\right]\nonumber\\
\pmb{A}_{BS}=&\left[\pmb{a}(M,\theta_{BS,1}),\pmb{a}(M,\theta_{BS,2}),\ldots,\pmb{a}(M,\theta_{BS,L})\right]&\nonumber\\
\pmb{\alpha}_k=&\textrm{diag}\left\{\alpha_1^{(k)},\alpha_2^{(k)},\ldots,\alpha_L^{(k)}\right\}\\
\pmb{D}_k=&\textrm{diag}\left\{e^{-j2\pi f\tau_1^{(k)}},e^{-j2\pi f\tau_2^{(k)}},e^{-j2\pi f\tau_L^{(k)}}\right\}\nonumber\\
\pmb{O}_{k,t}=&\textrm{diag}\left\{e^{j2\pi tf_{D,1}^{(k)}T_s},e^{j2\pi tf_{D,2}^{(k)}T_s},\ldots,e^{j2\pi tf_{D,L}^{(k)}T_s}\right\}\nonumber
\end{align}
As seen in \eqref{eq:ISAC-49} and \eqref{eq:ISAC-50}, only the Doppler-related terms in $\pmb{O}_{k,t}$ are time-dependent. 

To apply the compressed sensing\index{Sensing!compressed}
methods~\cite{5454399,9898900,9493736,10071555} for problem solving,
\eqref{eq:ISAC-49} can be further extended to use its virtual
representation~\cite{7400949}\index{Virtual representation!mmWave
  channel}. For this representation, the possible AoD range is divided
into $D>>L$ and also $D\geq M$ divisions corresponding to the angles
of $\{\theta_1,\theta_2,\ldots,\theta_D\}$, and the possible AoA range
is also divided into $D$ divisions corresponding to the angles of
$\{\phi_1,\phi_2,\ldots,\phi_D\}$. Then, two dictionaries are formed
as
\begin{align}\label{eq:ISAC-51}
\tilde{\pmb{A}}_u=&\left[\pmb{a}(N_k,\phi_{1}),\pmb{a}(N_k,\phi_{2}),\ldots,\pmb{a}(N_k,\phi_{D})\right]\nonumber\\
\tilde{\pmb{A}}_{BS}=&\left[\pmb{a}(M,\theta_{1}),\pmb{a}(M,\theta_{2}),\ldots,\pmb{a}(M,\theta_{D})\right]
\end{align}
where $\tilde{\pmb{A}}_u$ is the dictionary for users and
$\tilde{\pmb{A}}_{BS}$ is the dictionary used by BS.  With the aid of
these definitions, \eqref{eq:ISAC-49} can be virtually represented as
\begin{align}\label{eq:ISAC-52}
\pmb{H}_{k,t}=&\tilde{\pmb{A}}_u\tilde{\pmb{\Sigma}}_{k,t}\tilde{\pmb{A}}_{BS}^T\nonumber\\
=&\tilde{\pmb{A}}_u\tilde{\pmb{\alpha}}_k\tilde{\pmb{D}}_k\tilde{\pmb{O}}_{k,t}\tilde{\pmb{A}}_{BS}^T
\end{align} 
where $\tilde{\pmb{\Sigma}}_{k,t}$, and also $\tilde{\pmb{\alpha}}_k$,
$\tilde{\pmb{D}}_k$ and $\tilde{\pmb{O}}_{k,t}$ are all $(D\times D)$
diagonal matrices, with their values in \eqref{eq:ISAC-50} located at
the corresponding diagonal positions, where the angular divisions in
the dictionaries provide the closed approximations to the actual AoAs
in $\pmb{A}_k$ and the actual AoDs in $\pmb{A}_{BS}$ in
\eqref{eq:ISAC-50}.

Explicitly, $\textrm{vec}(\tilde{\pmb{\Sigma}}_{k,t})$ is a sparse
vector. Through sensing the positions and values of the non-zero
elements in $\textrm{vec}(\tilde{\pmb{\Sigma}}_{k,t})$, the variables,
including AoDs, AOAs, $\{\alpha_{l}^{(k)}\}$, $\{\tau_{l}^{(k)}\}$ and
$\{f_{D,l}^{(k)}\}$, can be derived, from which the position and
mobility behaviour of user $k$ can be determined. In more detail,
assume that through sparse recovery, the positions and values of the
non-zero elements in $\textrm{vec}(\tilde{\pmb{\Sigma}}_{k,t})$,
$t=1,2,\ldots,T$, generated by the $l$th path are obtained. Then, in
principle, first, based on the positions of the non-zero-elements, the
sensing receiver knows $\theta_{k,l}$ and $\phi_{k,l}$, from which
plus the locations of BS and user $k$ or plus the delay (to be
obtained below) from BS or from user $k$, the position of an object
generating the $l$th path can be determined. Note that, once
$\theta_{k,l}$ and $\phi_{k,l}$ for $l=1,\ldots,L$ are known, the
precoder and combiner can be derived. Second, the values of the
non-zero elements corresponding to the $l$th path over $T$
transmissions can be expressed as
\begin{align}\label{eq:ISAC-53}
\pmb{h}_l=&\left[\alpha_le^{-j2\pi f\tau_l}e^{j2\pi f_{D,l}T_s},\alpha_le^{-j2\pi f\tau_l}e^{j4\pi f_{D,l}T_s},\ldots,\alpha_le^{-j2\pi f\tau_l}e^{j2T\pi f_{D,l}T_s}\right]^T\nonumber\\
=&c_l\left[1,e^{j2\pi T_sf_{D,l}},\ldots,e^{j2\pi(T-1)T_s f_{D,l}}\right]^T
\end{align}
where $c_l=\alpha_le^{-j2\pi f\tau_l}e^{j2\pi f_{D,l}T_s}$ is an
unknown constant. When sampling $f_{D,l}$ to obtain
$f_{D,l}=f_x/(TT_s)$, where $f_x=0,1,\ldots,T-1$, \eqref{eq:ISAC-53}
becomes
\begin{align}\label{eq:ISAC-54}
\pmb{h}_l=&c_l\left[1,e^{j2\pi f_x/T},\ldots,e^{j2\pi(T-1)f_x/T}\right]^T
\end{align}
Explicitly, the part in bracket is the $f_x$th column of a $T$-point
IDFT matrix, as seen in \cite{Lie-Liang-MC-CDMA-book}. Therefore,
$f_x$, i.e., $f_{D,l}=f_x/(TT_s)$ can be derived via the FFT operation
on $\pmb{h}_l$, implied by the non-zero (or maximum magnitude) element
of $\pmb{\mathcal{F}}_T\pmb{h}_l$, where $\pmb{\mathcal{F}}_T$ is the
$T$-point FFT matrix.
   
After the estimation of $f_{D,l}=f_x$, we can express
$c_l=c_D\alpha_le^{-j2\pi f\tau_l}$, where $c_D$ is a constant
generated after removing the effect of Doppler-shift. Based on $c_l$,
we can obtain
\begin{align}\label{eq:ISAC-55}
&|\alpha_l|=|c_l|/c_D\nonumber\\
&2\pi f\tau_l+\varphi_l=-\arctan\left(\frac{\Im (c_l)}{\Re (c_l)}\right)
\end{align}
From the second equation of \eqref{eq:ISAC-55}, the delay can be
derived once the phase $\varphi_l$ associated with $\alpha_l$, which is
due to other effects, is known. Vice versa, if $\tau_l$ is known,
$\varphi_l$ can be estimated.

For example, when OFDM with $N$ subcarriers is
considered~\cite{9493736}, from $N$ subcarriers we will obtain a
vector
\begin{align}\label{eq:ISAC-56}
\pmb{c}_l=\tilde{\alpha}_l\left[1,e^{-j2\pi f\tau_l},\ldots,e^{-j2\pi (N-1)f\tau_l}\right]^T
\end{align}
after removing the effect of Doppler-shift, where $f$ represents the
subcarrier spacing and $\tilde{\alpha}_l$ is a modified version of
$\alpha_l$. Then, when sampling $\tau_l$ with a resolution of
$\Delta{\tau}=1/Nf$, we obtain $\tau_l=\tau_x/Nf$. Substituting it
into \eqref{eq:ISAC-56} gives
\begin{align}\label{eq:ISAC-57}
\pmb{c}_l=\tilde{\alpha}_l\left[1,e^{-j2\pi \tau_x/N},\ldots,e^{-j2\pi (N-1)\tau_x/N}\right]^T
\end{align}
where the bracket part is the $\tau_x$-column of a DFT matrix. Hence,
$\tau_x$ can be obtained from the IFFT operation on $\pmb{c}_l$,
giving the delay estimate to $\tau_l=\tau_x/Nf$. Furthermore, from the
IFFT operation, $\tilde{\alpha}_l$ can be obtained, from which the
original ${\alpha}_l$ can also be obtained after removing the effect
from the Doppler-shift. Consequently, the phase of $\alpha_l$, i.e.,
$\varphi_l$ in \eqref{eq:ISAC-55}, can be estimated. 

In summary, in OFDM mmWave systems, the estimations of AoD $\theta_l$,
AoA $\phi_l$, $\alpha_l$, $\tau_l$ and $f_{D,l}$ can be described
as Algorithm~\ref{Algorithm-OFDM-mmWave-par-estimation}

\begin{algorithm}[ht]
\caption{Algorithm for channel parameter estimation in OFDM mmWave
  systems.}
\label{Algorithm-OFDM-mmWave-par-estimation}

\textbf{Inputs:} receive observations $\pmb{y}_{k,t}$ or
$\pmb{y}_{s,t}$ $\forall t$ and subcarriers. \\

\textbf{Estimation of AoD and AoA}: Based on the virtual channel
representation of \eqref{eq:ISAC-52}, determine the AoDs
$\{\theta_l\}$ and AoAs $\{\phi_l\}$ with the aid of beam-search
algorithm~\cite{9493736}.\\

\textbf{Estimation of $f_{D,l}$, $\tau_l$ and $\alpha_l$}: For each
$l$, execute:\\

\begin{enumerate}

\item For each of subcarriers, estimate $f_{D,l}$ based on
  \eqref{eq:ISAC-54} by applying on it the $T$-point IDFT. The maximum
  terms corresponding to $N$ subcarriers generate a vector in the form
  of \eqref{eq:ISAC-56} and \eqref{eq:ISAC-57}.

\item Estimate $\tau_l$ by applying the $N$-point DFT on
  \eqref{eq:ISAC-57}.

\item Using any an element seen in \eqref{eq:ISAC-53} to estimate
  $\alpha_l=|\alpha|_le^{j\varphi_l}$.

\end{enumerate}

Alternatively, for each $l$, execute:\\
\begin{enumerate}

\item For each symbol $1\leq t\leq T$, estimate $\tau_{l}$ based on
  \eqref{eq:ISAC-57} by applying on it the $N$-point DFT. The maximum
  terms corresponding to $T$ symbols generate a vector in the form of
  \eqref{eq:ISAC-53} and \eqref{eq:ISAC-54}.

\item Estimate $f_{D,l}$ by applying the $T$-point IDFT on
  \eqref{eq:ISAC-54}.

\item Using any an element seen in \eqref{eq:ISAC-53} to estimate
  $\alpha_l=|\alpha|_le^{j\varphi_l}$.

\end{enumerate}

\end{algorithm}

Below several ISAC examples are analyzed with respect to resource-allocation. With the first one, we assume that there is one communication user. While BS sends information to the communication user, it also sends one to several beams to scan the environment for interesting targets. For communication, BS and user are assumed to know the communication channel $\pmb{H}_t$ in \eqref{eq:ISAC-42}, where the user index $k$ is ignored for simplicity. Hence, without considering the impact of sensing, BS and user are capable of designing the optimum or near-optimum precoder and combiner, which are expressed as $\pmb{F}_{c,t}\in\mathbb{C}^{M\times I_c}$ and $\pmb{W}_{c,t}\in\mathbb{C}^{N_c\times I_c}$, respectively, where $I_c$ is the number of data steams from BS to user and $N_c$ is the number of array elements of user's receive array. 

Assume that, in each time interval of certain duration, BS also sends
$I_s$ beams in $I_s$ different directions to sense possible
objects. For designing the sensing beams, when BS knows $\pmb{Q}_h$ of
channels' correlation matrices, the beams may be designed based on the
method described in Section~\ref{subsection-6G-6.2.2}. Otherwise, when
without considering the second-order statistics of channels, the $I_s$
scanning beans can be designed by solving the
equation~\cite{book:Van-Trees}
\begin{align}\label{eq:ISAC-58}
\tilde{\pmb{A}}_{BS}^T\pmb{F}_s=\pmb{G}
\end{align}        
where $\tilde{\pmb{A}}_{BS}\in\mathbb{C}^{M\times D}$ is given in \eqref{eq:ISAC-51}, $\pmb{F}_s\in\mathbb{C}^{M\times I_s}$ is to be found, and $\pmb{G}\in\mathbb{C}^{D\times I_s}$ defines the desired array responses in the individual directions determined by the vectors in $\tilde{\pmb{A}}_{BS}$. For example, we may simply set $\pmb{G}$ to the first $I_s$ columns of identity matrix $\pmb{I}_D$. Since $D\geq M$, the solutions to $\pmb{F}_s$ can be simply found in zero-forcing (ZF)\index{Zero-forcing (ZF)} principle as
\begin{align}\label{eq:ISAC-59}
\pmb{F}_s=\left(\tilde{\pmb{A}}_{BS}^*\tilde{\pmb{A}}_{BS}^T\right)^{-1}\tilde{\pmb{A}}_{BS}^*\pmb{G}
\end{align}        

However, $\pmb{F}_s$ can only scan $I_s$ directions, but BS may want
to scan many more directions using different time intervals. To
achieve this, let us re-visit the array response vector in
\eqref{eq:ISAC-47} and assume $\theta=\theta_1+\theta_2$, yielding
$\vartheta=\vartheta_1+\vartheta_2$. Then, \eqref{eq:ISAC-47} can have
the form of
\begin{align}\label{eq:ISAC-60}
\pmb{a}(N,\theta_1+\theta_2)=&\left[1,e^{-j2\pi(\vartheta_1+\vartheta_2)},e^{-j4\pi(\vartheta_1+\vartheta_2)},\ldots,e^{-j2(N-1)\pi(\vartheta_1+\vartheta_2)}\right]^T/\sqrt{N}\nonumber\\
=&\textrm{diag}\{\sqrt{N}\pmb{a}(N,\theta_1)\}\pmb{a}(N,\theta_2)
\end{align}
which shows that a new directional vector $\pmb{a}(N,\theta_1+\theta_2)$ can be obtained by multiplying a diagonal mapping matrix $\textrm{diag}\{\sqrt{N}\pmb{a}(N,\theta_1)\}$ on the old directional vector $\pmb{a}(N,\theta_1)$. 

Hence, assume that the total $I$ directions need $J=I/s$ time
intervals to complete their scan. First, an initial $\pmb{F}_s$ and a
scanning interval $\Delta\vartheta$ can be appropriately
designed. Then, the precoding matrices for the $J$ intervals can be
formed as
$\pmb{F}_s(j)=\textrm{diag}\left\{\sqrt{N}\pmb{a}\left(N,\arcsin\left(\frac{\lambda
  j\Delta\vartheta}{d}\right)\right)\right\}\pmb{F}_s$ for
$j=0,1,\ldots,J-1$.

Even without considering ISAC, the above precoder design for
communication or sensing has the implication of power-allocation under
the constraint on power. Specifically, as analyzed in Section 8.3.5 of
\cite{Lie-Liang-MC-CDMA-book}, if the individual power normalization
is applied, each column of $\pmb{F}_{c,t}$ and $\pmb{F}_s(j)$ should
be normalized to unit length. On the other side, when the total
transmit power is constraint, $\pmb{F}_{c,t}$ and $\pmb{F}_s(j)$ need
to be constrained by $\textrm{Tr}(\pmb{F}_{c,t}^H\pmb{F}_{c,t})\leq
I_c$ and $\textrm{Tr}(\pmb{F}^H_s(j)\pmb{F}_s(j))\leq I_s$.
Furthermore, the power to different beams may be assigned to maximize
SINR or mutual information (See Sections 8.2.6 and 8.2.11 of
\cite{Lie-Liang-MC-CDMA-book}, respectively).

The power-allocation in ISAC mmWave system becomes more challenging. For example, assume that $\pmb{F}_{c,t}$ and  $\pmb{F}_s(j)$ have been appropriately normalized. Now the ISAC mmWave system motivates to design an ISAC precoder for jointly transmitting information to a user and scanning all the interested directions, as above-mentioned, in $J$ time intervals, by combining $\pmb{F}_{c,t}$ and $\pmb{F}_s(j)$ designed without considering their inter-effect~\cite{8550811,8738892,9131843}. Let the information symbol vectors are $\pmb{s}_{c,t}\in\mathbb{C}^{I_c\times 1}$ and probing signal vectors are $\pmb{s}_{s,t}\in\mathbb{C}^{I_s\times 1}$. The simplest method to form the ISAC transmit signals is 
\begin{align}\label{eq:ISAC-61}
\pmb{x}_t=\sqrt{\rho}\pmb{F}_{c,t}\pmb{s}_{c,t}+\sqrt{1-\rho}\pmb{F}_s(j)\pmb{s}_{s,t}
\end{align}
where the coefficient, $0\leq \rho\leq 1$, explains the power assigned
between communication and sensing, which needs to be optimized to
maximize (or guarantee) communication performance while guarantee (or
maximize) the sensing performance. This method is relatively low
complexity, but degrades the communication and sensing performance due
to their interference.

To improve performance at the cost of some increased complexity, \eqref{eq:ISAC-61} can be enhanced to
\begin{align}\label{eq:ISAC-62}
\pmb{x}_t=\sqrt{\rho}\pmb{F}_{c,t}\pmb{s}_{c,t}+\sqrt{1-\rho}\pmb{F}_s(j)\pmb{\beta}\pmb{s}_{s,t}
\end{align}
to focus on the optimization of communication performance. In \eqref{eq:ISAC-62},  $\pmb{\beta}\in\mathbb{C}^{I_s\times I_s}$ is a diagonal matrix. The elements of $\pmb{\beta}$ and parameter $\rho$ can be optimized under the constraint of $\sum_{i=1}^{I_s}\|\beta_{ii}\|^2=I_s$, with the objective to, such as, maximize the SINR of communication receiver. Notice in \eqref{eq:ISAC-62} that the precoder for sensing is modified to $\pmb{F}_s(j)\pmb{\beta}$, meaning that a column is amplified by a magnitude $|\beta_{ii}|$ and shifted by a phase of $\arg(\beta_{ii})$, which may slightly change the beam patterns of sensing, but usually insignificant~\cite{8550811,8738892,9131843}. \eqref{eq:ISAC-61} can also be modified to
\begin{align}\label{eq:ISAC-63}
\pmb{x}_t=\sqrt{\rho}\pmb{F}_{c,t}\pmb{\beta}\pmb{s}_{c,t}+\sqrt{1-\rho}\pmb{F}_s(j)\pmb{s}_{s,t}
\end{align}
to focus on the optimization of sensing performance. In this case, the
elements of $\pmb{\beta}\in\mathbb{C}^{I_c\times I_c}$ and parameter
$\rho$ can be optimized to maximize the sensing
performance. Furthermore, the side effect on communication from
optimization may be mitigated by the combiner design at the
communication user.

The above optimization and processing with respect to $\pmb{\beta}$
have to be implemented in baseband, which demands relatively higher
complexity, especially when the mmWave systems with a huge number of
antennas are considered. To reduce implementation complexity, we may
only implement analog processing in RF domain by only optimizing the
phases of radiations via setting
$\pmb{\beta}=\textrm{diag}\{e^{j\phi_1},e^{j\phi_2},\ldots,e^{j\phi_{I_s}}\}$
in \eqref{eq:ISAC-62}, or
$\pmb{\beta}=\textrm{diag}\{e^{j\phi_1},e^{j\phi_2},\ldots,e^{j\phi_{I_c}}\}$
in \eqref{eq:ISAC-63}. Note that, applying this $\pmb{\beta}$ only
shifts the phases of the elements of a precoding vector in $\pmb{F}$
by a common phase, which can be achieved by multiplying this common
phase with all the elements of the transmit array.

In \eqref{eq:ISAC-61}-\eqref{eq:ISAC-63}, it is assumed that
independent probing signals are used for sensing. For the downlink
active sensing where BS senses for objects, BS knows ideally the
information sent to communication user. Hence, a signal sent to a
downlink user can be simultaneously used for sensing the objects in
environment. With this regard, we set
$\pmb{s}_{s,t}=\pmb{s}_{c,t}$. After BS receives the echo signals, it
can first remove $\pmb{s}_{c,t}$ and then execute the sensing
processing. By contrast, when the downlink passive sensing, where
communication user senses for objects, is considered, communication
user does not have the {\em a-priori} information of $\pmb{s}_{c,t}$,
it has to implement the so-called decision-directed
sensing\index{Decision-directed sensing}. In other words, it first
detects $\pmb{s}_{c,t}$ and then uses the detected symbols to cancel
the communication information, followed by the sensing
processing. Note that this kind of ISAC design focuses more on the
optimization of communication to maximize the communication
performance, while under the constraint of sensing performance.

Specifically, assume one communication user with one data stream and one scanning beam~\cite{8550811,8738892,9131843}, the transmitted ISAC signal can be written as  
\begin{align}\label{eq:ISAC-64}
\pmb{x}_t=\left[\sqrt{\rho}\pmb{f}_{c,t}+\sqrt{1-\rho}e^{j\phi}\pmb{f}_s(j)\right]\pmb{s}_{c,t}
\end{align}
where $\pmb{f}_{c,t}$ and $\pmb{f}_s(j)$ are the precoding vectors for
communication and sensing, respectively, which are designed without
considering their inter-effect. In \eqref{eq:ISAC-64}, the parameters
$\rho$ and $\phi$ can be optimized separately. $\rho$ can be
determined according to the communication and sensing ranges. The
optimization in the context of $\phi$ ensures that the signals for
communication and sensing can be coherently combined at communication
receiver, and hence enhances the communication performance. In
\cite{8550811,8738892,9131843}, various algorithms have been proposed
to achieve the objective.

Finally, it is worth noting that ISAC signals can be designed jointly, as discussed in Section~\ref{subsection-6G-6.3} and as seen, for example, in \cite{8999605,9540344,9898900,8288677} and some references there in. Also, the precoders can be designed to make the communication and sensing signals fall in the orthogonal spaces and hence, they do not interfere with each other, as seen in Chapter 8 of \cite{Lie-Liang-MC-CDMA-book}. Additionally, we should note that the above special cases only consider one communication user for simplicity. Its extension to the ISAC mmWave systems supporting multiple downlink communications users is straightforward. In this case, the signal processing for precoder/combiner design as well as the optimization of resource-allocation will undoubtedly become more challenging.     

\index{ISAC!mmWave|)}

%%%%%%%%%%
\section{Concluding Remarks}
%%%%%%%%%%

The fundamentals of MIMO communications and MIMO sensing were firstly
analyzed, demonstrating that communication and sensing have different
objectives, which result in that the signaling waveforms required to
achieve efficient communication and sensing are different. Then,
several design and resource optimization issues in the context of the
MIMO ISAC systems were discussed. Furthermore, the ISAC design and
optimization in mmWave ISAC systems were explored, where several
exemplified cases were specifically considered and a range of
optimization methods were reviewed.

It can be shown that, while the optimum design of ISAC is challenging,
communication and sensing are capable of benefiting each other,
especially, in mmWave communications. Sensing uses probing beams to
identify objects and communication terminals, which can be exploited
to significantly reduce the overhead and time required by the beam
search process in mmWave communications. Sensing can also be
implemented to enhance the channel estimation for improving the
performance of communications. On the other side, communication
signals can directly be exploited for object/environment sensing, in
addition to the positioning of communication terminals
themselves. With the aid of the embedded sensing, future wireless
communications can be expected to be carried out with the detailed
knowledge about any possible propagation paths/rays in an
environment. Hence, signal transmissions can be optimized on the basis
of individual propagation paths/rays, which enables path/ray
diversity, providing the information transmission of extremely high
space-efficiency\index{Space-efficiency}\footnote{Space-efficiency can
  be defined as the number of bits conveyed per second per Hertz over
  a unit of space.} and energy-efficiency.

Now that ISAC can only optimize the trade-off between the performance
of communication and that of sensing, in practice, the optimization of
communication may need to be emphasised in slow time-varying
environments, and otherwise, the optimization of sensing needs to be
focused in fast time-varying environments. This is because in slow
time-varying environments, the objects being sensed are nearly
stationary for a relatively long time, there may be a plenty of time
for collecting sufficient data to obtain reliable sensing. By
contrast, when in fast time-varying environments, reliable sensing is
desired to be achieved based on short-time observations, which is
undoubtedly challenging. Moreover, when operated in fast time-varying
environments, the performance of communications is crucially relied on
the availability of instantaneous CSI, which sensing can help and, in
turn, explains the importance of reliable sensing.

%\bibliography{Resource-Allocation}

\begin{thebibliography}{10}
\providecommand{\url}[1]{#1}
\csname url@samestyle\endcsname
\providecommand{\newblock}{\relax}
\providecommand{\bibinfo}[2]{#2}
\providecommand{\BIBentrySTDinterwordspacing}{\spaceskip=0pt\relax}
\providecommand{\BIBentryALTinterwordstretchfactor}{4}
\providecommand{\BIBentryALTinterwordspacing}{\spaceskip=\fontdimen2\font plus
\BIBentryALTinterwordstretchfactor\fontdimen3\font minus
  \fontdimen4\font\relax}
\providecommand{\BIBforeignlanguage}[2]{{%
\expandafter\ifx\csname l@#1\endcsname\relax
\typeout{** WARNING: IEEEtran.bst: No hyphenation pattern has been}%
\typeout{** loaded for the language `#1'. Using the pattern for}%
\typeout{** the default language instead.}%
\else
\language=\csname l@#1\endcsname
\fi
#2}}
\providecommand{\BIBdecl}{\relax}
\BIBdecl

\bibitem{9737357}
F.~Liu, Y.~Cui, C.~Masouros, J.~Xu, T.~X. Han, Y.~C. Eldar, and S.~Buzzi,
  ``Integrated sensing and communications: Toward dual-functional wireless
  networks for 6g and beyond,'' \emph{IEEE Journal on Selected Areas in
  Communications}, vol.~40, no.~6, pp. 1728--1767, 2022.

\bibitem{8827589}
M.~L. Rahman, J.~A. Zhang, X.~Huang, Y.~J. Guo, and R.~W. Heath, ``Framework
  for a perceptive mobile network using joint communication and radar
  sensing,'' \emph{IEEE Transactions on Aerospace and Electronic Systems},
  vol.~56, no.~3, pp. 1926--1941, 2020.

\bibitem{book:Thomas-Cover-Information-Theory}
T.~M. Cover and J.~A. Thomas, \emph{Elements of Information Theory}.\hskip 1em
  plus 0.5em minus 0.4em\relax New York: John Wiley \& Sons, 1991.

\bibitem{Lie-Liang-MC-CDMA-book}
L.-L. Yang, \emph{Multicarrier Communications}.\hskip 1em plus 0.5em minus
  0.4em\relax Chichester, United Kingdom: John Wiley, 2009.

\bibitem{book-Steven-Kay-I}
S.~M. Kay, \emph{Fundamentals of Statistical Signal Processing: Estimation
  Theory}.\hskip 1em plus 0.5em minus 0.4em\relax Upper Saddle River, New
  Jersey: Prentice Hall, Inc., 1993.

\bibitem{book:Narayan-Giri}
N.~C. Giri, \emph{Multivariate Statistical Analysis}, 2nd~ed.\hskip 1em plus
  0.5em minus 0.4em\relax New York: Marcel Dekker, Inc, 2004.

\bibitem{7279172}
A.~R. Chiriyath, B.~Paul, G.~M. Jacyna, and D.~W. Bliss, ``Inner bounds on
  performance of radar and communications co-existence,'' \emph{IEEE
  Transactions on Signal Processing}, vol.~64, no.~2, pp. 464--474, 2016.

\bibitem{5467189}
B.~Tang, J.~Tang, and Y.~Peng, ``{MIMO} radar waveform design in colored noise
  based on information theory,'' \emph{IEEE Transactions on Signal Processing},
  vol.~58, no.~9, pp. 4684--4697, 2010.

\bibitem{MIMO-Telatar-I}
I.~E. Telatar, ``Capacity of multiantenna {G}aussian channels,'' \emph{European
  Transactions on Telecommunications}, vol.~10, no.~6, pp. 585--595, Nov./Dec.
  1999.

\bibitem{9303435}
X.~Yuan, Z.~Feng, J.~A. Zhang, W.~Ni, R.~P. Liu, Z.~Wei, and C.~Xu,
  ``Spatio-temporal power optimization for {MIMO} joint communication and radio
  sensing systems with training overhead,'' \emph{IEEE Transactions on
  Vehicular Technology}, vol.~70, no.~1, pp. 514--528, 2021.

\bibitem{7970102}
Y.~Liu, G.~Liao, J.~Xu, Z.~Yang, and Y.~Zhang, ``Adaptive {OFDM} integrated
  radar and communications waveform design based on information theory,''
  \emph{IEEE Communications Letters}, vol.~21, no.~10, pp. 2174--2177, 2017.

\bibitem{9540344}
J.~A. Zhang, F.~Liu, C.~Masouros, R.~W. Heath, Z.~Feng, L.~Zheng, and
  A.~Petropulu, ``An overview of signal processing techniques for joint
  communication and radar sensing,'' \emph{IEEE Journal of Selected Topics in
  Signal Processing}, vol.~15, no.~6, pp. 1295--1315, 2021.

\bibitem{8386661}
F.~Liu, L.~Zhou, C.~Masouros, A.~Li, W.~Luo, and A.~Petropulu, ``Toward
  dual-functional radar-communication systems: Optimal waveform design,''
  \emph{IEEE Transactions on Signal Processing}, vol.~66, no.~16, pp.
  4264--4279, 2018.

\bibitem{8999605}
F.~Liu, C.~Masouros, A.~P. Petropulu, H.~Griffiths, and L.~Hanzo, ``Joint radar
  and communication design: Applications, state-of-the-art, and the road
  ahead,'' \emph{IEEE Transactions on Communications}, vol.~68, no.~6, pp.
  3834--3862, 2020.

\bibitem{8550811}
J.~A. Zhang, X.~Huang, Y.~J. Guo, J.~Yuan, and R.~W. Heath, ``Multibeam for
  joint communication and radar sensing using steerable analog antenna
  arrays,'' \emph{IEEE Transactions on Vehicular Technology}, vol.~68, no.~1,
  pp. 671--685, 2019.

\bibitem{9131843}
Y.~Luo, J.~A. Zhang, X.~Huang, W.~Ni, and J.~Pan, ``Multibeam optimization for
  joint communication and radio sensing using analog antenna arrays,''
  \emph{IEEE Transactions on Vehicular Technology}, vol.~69, no.~10, pp.
  11\,000--11\,013, 2020.

\bibitem{8288677}
F.~Liu, C.~Masouros, A.~Li, H.~Sun, and L.~Hanzo, ``{MU-MIMO} communications
  with {MIMO} radar: From co-existence to joint transmission,'' \emph{IEEE
  Transactions on Wireless Communications}, vol.~17, no.~4, pp. 2755--2770,
  2018.

\bibitem{9898900}
Z.~Gao, Z.~Wan, D.~Zheng, S.~Tan, C.~Masouros, D.~W.~K. Ng, and S.~Chen,
  ``Integrated sensing and communication with {mmWave} massive {MIMO}: A
  compressed sampling perspective,'' \emph{IEEE Transactions on Wireless
  Communications}, vol.~22, no.~3, pp. 1745--1762, 2023.

\bibitem{10275023}
Y.~Cui, H.~Ding, S.~Ke, and L.~Zhao, ``Integrated sensing and communication in
  {mmWave} wireless backhaul networks,'' \emph{IEEE Transactions on Vehicular
  Technology}, pp. 1--15, 2023.

\bibitem{10316594}
J.~Zhang, S.~Yan, and M.~Peng, ``Joint beam alignment and resource allocation
  for multi-user {mmWave} integrated sensing and communication systems,''
  \emph{IEEE Transactions on Vehicular Technology}, pp. 1--16, 2023.

\bibitem{7400949}
R.~W. Heath, N.~Gonzalez-Prelcic, S.~Rangan, W.~Roh, and A.~M. Sayeed, ``An
  overview of signal processing techniques for millimeter wave {MIMO}
  systems,'' \emph{IEEE Journal of Selected Topics in Signal Processing},
  vol.~10, no.~3, pp. 436--453, 2016.

\bibitem{5454399}
W.~U. Bajwa, J.~Haupt, A.~M. Sayeed, and R.~Nowak, ``Compressed channel
  sensing: A new approach to estimating sparse multipath channels,''
  \emph{Proceedings of the IEEE}, vol.~98, no.~6, pp. 1058--1076, 2010.

\bibitem{9493736}
K.~Li, M.~El-Hajjar, and L.-l. Yang, ``Millimeter-wave based localization using
  a two-stage channel estimation relying on few-bit {ADCs},'' \emph{IEEE Open
  Journal of the Communications Society}, vol.~2, pp. 1736--1752, 2021.

\bibitem{10071555}
K.~Li, M.~El-Hajjar, and L.-L. Yang, ``Reconfigurable intelligent surface aided
  position and orientation estimation based on joint beamforming with limited
  feedback,'' \emph{IEEE Open Journal of the Communications Society}, vol.~4,
  pp. 748--767, 2023.

\bibitem{book:Van-Trees}
H.~L.~V. Trees, \emph{Optimum Array Processing}.\hskip 1em plus 0.5em minus
  0.4em\relax Wiley Interscience, 2002.

\bibitem{8738892}
Y.~Luo, J.~A. Zhang, X.~Huang, W.~Ni, and J.~Pan, ``Optimization and
  quantization of multibeam beamforming vector for joint communication and
  radio sensing,'' \emph{IEEE Transactions on Communications}, vol.~67, no.~9,
  pp. 6468--6482, 2019.

\end{thebibliography}


\end{document}